\documentclass[a4paper,twocolumn,showpacs,superscriptaddress]{revtex4}

\usepackage[english]{babel}
\usepackage{graphicx}
\usepackage{xspace}
\usepackage{amsmath,amssymb}

\DeclareMathAlphabet{\mathbfit}{OT1}{cmr}{bx}{it}

\begin{document}

\title{Superconductivity phase diagrams of electron doped cuprates
$R_{2-x}$Ce$_x$CuO$_4$ ($R=$\,La, Pr, Nd, Sm, and Eu)}

\author{Y.~Krockenberger}\altaffiliation{e-mail:
yoshiharu.krockenberger@aist.go.jp} \affiliation{Institute of
Materials Science, TU Darmstadt, Petersenstr.~23, 64287 Darmstadt,
Germany}\affiliation{NTT Basic Research Laboratories, 3-1
Wakamiya, Atsugi-shi, Kanagawa 243-0198, Japan}
\author{J.~Kurian}
\affiliation{Institute of Materials Science, TU Darmstadt,
Petersenstr.~23, 64287 Darmstadt, Germany}\affiliation{NTT Basic
Research Laboratories, 3-1 Wakamiya, Atsugi-shi, Kanagawa
243-0198, Japan}
\author{A.~Winkler}
\affiliation{Institute of Materials Science, TU Darmstadt,
Petersenstr.~23, 64287 Darmstadt, Germany}\affiliation{NTT Basic
Research Laboratories, 3-1 Wakamiya, Atsugi-shi, Kanagawa 243-0198,
Japan}
\author{A.~Tsukada}
\affiliation{NTT Basic Research Laboratories, 3-1 Wakamiya,
Atsugi-shi, Kanagawa 243-0198, Japan} \affiliation{Department of
Applied Physics, Tokyo University of Agriculture and Technology
(TUAT), Japan}
\author{M.~Naito}
\affiliation{NTT Basic Research Laboratories, 3-1 Wakamiya,
Atsugi-shi, Kanagawa 243-0198, Japan} \affiliation{Department of
Applied Physics, Tokyo University of Agriculture and Technology
(TUAT), Japan}
\author{L.~Alff}
\email{alff@oxide.tu-darmstadt.de} \affiliation{Institute of
Materials Science, TU Darmstadt, Petersenstr.~23, 64287 Darmstadt,
Germany}

\date{9th December 2007}

\begin{abstract}
The superconductivity phase diagrams of electron doped cuprates of
the form $R_{2-x}$Ce$_x$CuO$_4$ (with $R=$\,La, Pr, Nd, Sm, and
Eu) have been determined for cerium compositions $0 < x < 0.36$ in
a consistent series of epitaxial thin films grown by reactive
molecular beam epitaxy (MBE). The use of epitaxial thin films
allows the growth of materials away from thermodynamical
equilibrium expanding the accessible phase space beyond the
availability of bulk material. The superconducting phase space
systematically increases with the rare earth ionic size. The
doping concentration where the maximal transition temperature
occurs in La$_{2-x}$Ce$_x$CuO$_4$ is considerably shifted to lower
doping ($x\sim0.09$) compared to La$_{2-x}$Sr$_x$CuO$_4$
($x\sim0.15$). At the same time, the width of the superconducting
region is broadened.
\end{abstract}
\maketitle

The phase diagram of cuprate superconductors is a key ingredient
to understand the still unresolved mechanism of high-temperature
superconductivity. A particular interesting question is the
comparison of the phase diagrams for hole and electron doping
\cite{Tokura:89}. A theory of high-temperature superconductivity
has to explain the occuring differences and similarities when a
copper-oxide plane is doped by holes or electrons. While for the
hole doped case there is already an overwhelming amount of
experimental data available, the electron doped side of the phase
diagram still needs {\em experimental clarification}.

Electron doped cuprates are widely identified with the two
materials Nd$_{2-x}$Ce$_x$CuO$_4$ \cite{Luke:90} and
Pr$_{2-x}$Ce$_x$CuO$_4$. In contrast to the hole doped cuprates
where in compounds of the form $R_{2-x}$Sr$_x$CuO$_4$ only for
$R=$ La superconductivity shows up, for the electron doped side of
the phase diagram a whole family of superconductors of the form
$R_{2-x}$Ce$_x$CuO$_4$ ($LN=$ La, Pr, Nd, Sm, and Eu) exists. It
is solely due to historical reasons that most investigations of
electron-doped cuprates have been made for Nd$_{2-x}$Ce$_x$CuO$_4$
and Pr$_{2-x}$Ce$_x$CuO$_4$. These two materials have an almost
identical ionic radius and show, therefore, also a very similar
phase diagram. This has led to the common belief that this
properties of a {\em specific} electron doping phase diagram (in
particular a narrow superconducting region) is intrinsic to
electron doped cuprates {\em in general}. In addition, for
Nd$_{2-x}$Ce$_x$CuO$_4$ (and to a lesser extent for
Pr$_{2-x}$Ce$_x$CuO$_4$) the magnetism of Nd$^{3+}$ (and
Pr$^{3+}$) masks in some experiments the intrinsic properties of
the superconductor like the temperature dependence of the London
penetration depth \cite{Alff:99}. Moreover, the difficulties in
bulk sample preparation makes the determination of exact phase
diagrams a much debated topic \cite{Kang:03,Mang:03,Kang:03a}

Using reactive MBE  \cite{Naito:00} and pulsed laser deposition
(PLD) \cite{Sawa:02} recently superconducting epitaxial thin films
of La$_{2-x}$Ce$_x$CuO$_4$ have been successfully grown. For bulk
material up to now only the mixed compound
LaPr$_{1-x}$Ce$_x$CuO$_4$ could be synthesized
\cite{Uefuji:02,Fujita:03}, and is now intensively studied
\cite{Matsui:05,Wilson:06,Wilson:06n}. Here we present a new
consistent and detailed study of the complete electron doped
family $R_{2-x}$Ce$_x$CuO$_4$ (with $R=$\,La, Pr, Nd, Sm, Eu, and
Gd) based on MBE-grown epitaxial thin films. The key results are
that the width of the superconducting phase space increases
systematically with the ionic size of the rare earth element $LN$,
and that the doping level of maximal superconductivity with
highest critical temperature, $T_{\text{C}}$, is not fixed (to
$x\approx0.15$), but shifts to significantly lower doping.

We conclude that the most meaningful comparison of the
superconductivity phase diagrams of hole and electron doped
cuprates, that can be done given this material's situation, should
be made using La-based cuprates. However, one still has to keep in
mind that La$_{2-x}$Sr$_x$CuO$_4$ has the so-called $T$-structure
(K$_2$NiF$_4$ structure) containing apical oxygen, while
La$_{2-x}$Ce$_x$CuO$_4$ with $x\gtrsim0.05$ has the so-called
$T'$-structure (Nd$_2$CuO$_4$ structure) where the Cu ion is only
fourfold coordinated, i.e., the apical oxygen ions are missing.
The structural difference translates into different ionic radii of
the Cu$^{2+}$ ions: $0.73$\,{\AA} in the $T$-structure, and
$0.57$\,{\AA} in the $T'$-structure (reduction of more than 20\%).
It is still one of the most exciting future challenges in
high-$T_{\text{C}}$ research to find a system, where a {\em
direct} comparison of hole and electron doping can be made in a
wide range of doping.
\begin{figure}[t]
\centering{%
\includegraphics[width=0.9\columnwidth,clip=]{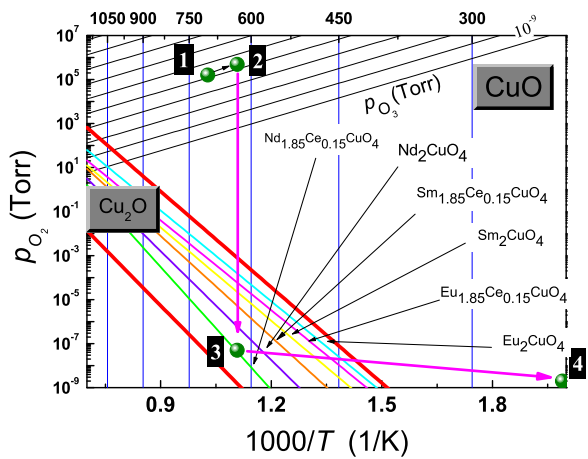}}
 \caption{(Color online) Typical thermodynamic phase stability diagram for electron doped
cuprates. Stability lines for CuO and Cu$_{2}$O have been
calculated using the commercially available program
\texttt{MALT}$^{\circledR}$. Additionally, the equilibria
oxidizing potential lines for ozone and oxygen are calculated. The
numbered points describe a typical growth of the thin film. The
border lines for the different copper valencies are thick red
colored. Between them, all experimentally established stability
lines for the $T^{\prime}$-structure compounds are lying. Point 1
and 2 represents the growth followed by annealing in vacuum (point
3) and afterwards cold down to point 4.}\label{Fig:syn}
\end{figure}
All our films were grown by reactive MBE using simultaneous
electron-beam evaporation from elemental sources on (001)
SrTiO$_3$ substrates. The stoichiometry was controlled by
electron-impact emission spectroscopy (EIES). A typical growth
scenario of electron doped superconducting thin films is shown in
Fig.~\ref{Fig:syn}. Films are grown at a substrate temperature of
about 700$^\circ$C in an ozone pressure of about
$2\cdot10^{-6}$\,Torr. During growth the sample has to be in the
region of divalent copper (correspondingly the rare earth elements
are $3+$, and Ce $4+$). Growth rates are approximately 3\,{\AA}/s
leading to homogeneous and precipitate free epitaxial thin films.

It is well known that the oxygen contents of electron doped
cuprates is a crucial and so far not fully understood issue
\cite{Richard:04,Higgins:06,Yu:07,Kang:07}. We have adopted a
consistent sample treatment in order to exclude as far as possible
different oxygen contents in our samples: all samples were
annealed in vacuum at temperatures close to the stability line of
each compound. The stability lines shown in Fig.~\ref{Fig:syn}
have been determined experimentally by observing in real time the
reflection high-energy electron diffraction (RHEED) signal which
indicates starting decomposition. This procedure adopted in the
present investigation enables the comprehensive removal of excess
oxygen without any measurable decomposition due to too strong
reduction. In the case of bulk material synthesis, the reduction
process may either lead to non-complete removal of excess oxygen,
or to partial decomposition of the material
\cite{Kang:07,Hauck:98}. Another important issue for electron
doped cuprates is the presence of disorder that can affect
$T_{\text{C}}$ seriously. Taking room-temperature resistivity as a
measure of disorder, for our study a correlation between disorder
and ionic radii of the lanthanide ions can be excluded, since all
compounds $R_{2-x}$Ce$_x$CuO$_4$ (with $R=$\,La, Pr, Nd, and Sm)
have at optimal doping very similar resistivity, except for Eu
where the possible role of disorder cannot be ruled out. For all
samples Ce-doping reduces linearly the $c$-axis parameter, because
the ionic radius of Ce$^{4+}$ is $0.97$\,{\AA} which is smaller
than the radius of the rare earth elements. The typical thickness
of the films in the present study was $\sim$1000\,\AA.

{\em La$_{2-x}$Ce$_x$CuO$_4$}: It would be of course a breakthrough
in the research on electron-doped cuprates to make available bulk
material of La$_{2-x}$Ce$_x$CuO$_4$ in the Nd$_2$CuO$_4$ structure.
For thin film growth in ultra high vacuum far from thermodynamic
equilibrium, epitaxial growth of $T'$-La$_{2-x}$Ce$_x$CuO$_4$ is
possible at a substrate temperature of 700$^\circ$C and in a partial
ozone pressure of $2\times10^{-6}$\,Torr. As has been shown before,
the structural phase transition into the $T$-phase occurs for
$x\lesssim0.05$ \cite{Naito:00}. However, a further reduction of
growth temperature by 100$^\circ$C allows the growth of $T'$-phase
La$_{2-x}$Ce$_x$CuO$_4$ even down to $x=0$ - but at the cost of
crystalline quality. It is even possible to stabilize the $T'$-phase
with improved crystallinity by substitution of smaller trivalent
ions like Tb, Y etc.~\cite{Tsukada:05}.

\begin{figure}[t]
\centering{%
\includegraphics[width=0.85\columnwidth,clip=]{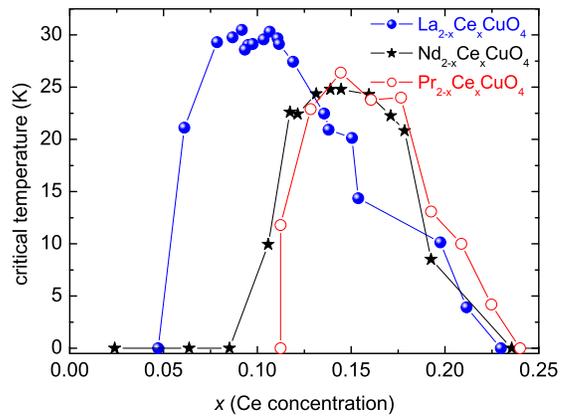}}
 \caption{(Color online) Superconductivity phase diagrams of La$_{2-x}$Ce$_x$CuO$_4$,
 Nd$_{2-x}$Ce$_x$CuO$_4$, and Pr$_{2-x}$Ce$_x$CuO$_4$.}\label{Fig:LCCO}
\end{figure}

In Fig.~\ref{Fig:LCCO} we show the obtained phase diagram for
La$_{2-x}$Ce$_x$CuO$_4$. The two major observations are (i) the
strongly broadened superconductivity region ranging from
$x\approx0.05$ to $x\approx0.22$, and (ii) the maximal
$T_{\text{C}}$ of 32\,K occuring at $x\approx0.09$. Sawa {\em et
al.}~\cite{Sawa:02} claim a complete shift (i.e., no broadening)
of the superconductivity region based on thin films grown by PLD
if samples have high dopant homogeneity. However, the broader
superconducting region as confirmed in this paper is clearly not
due to any inhomogeneities of the samples. First, the sample size
is too small (3\,mm by 5\,mm) to expect compositional
inhomogeneities in the given MBE setup. For example, three
different samples attached to the heater and grown in the same run
give exactly the same $T_{\text{C}}$. Second, the resistivity
values of the MBE grown films are well below those of the PLD thin
films \cite{Naito:00,Sawa:02}. Third, the critical temperature in
the MBE grown thin films (see Fig.~\ref{Fig:LCCO}) is higher than
compared to PLD samples \cite{Sawa:02}. Fourth, the
superconducting transition width in resistivity and magnetometry
is even sharper for higher doping where the extended
superconducting region is observed in this study. The room
temperature resistivity values of La$_{2-x}$Ce$_x$CuO$_4$ films
lies between 0.22--2.0\,m$\Omega$cm for optimal- and undoped
samples, respectively.

{\em Pr$_{2-x}$Ce$_x$CuO$_4$ and Nd$_{2-x}$Ce$_x$CuO$_4$}: The
corresponding superconductivity phase diagrams of MBE grown films
are included in Fig.~\ref{Fig:LCCO} for a direct comparison. The
superconductivity region extends from $x\approx0.10$ to
$x\approx0.24$, and the maximal $T_{\text{C}}$ of $26$\,K occurs
for $x\approx0.145$. Note also, that UHV-annealing times and
growth temperatures differ slightly for different $x$. In this
study, we have always used the parameters yielding the highest
$T_{\text{C}}$ for a given doping concentration $x$. The simple
trend is, that for the highest crystallinity (as indicated in a
standard x-ray diffraction pattern by the intensity of the (006)
reflection of the epitaxial thin film), also the highest
$T_{\text{C}}$ is obtained. The phase diagram based on thin film
data fully agrees with the huge amount of published data for bulk
material of Pr$_{2-x}$Ce$_x$CuO$_4$ and Nd$_{2-x}$Ce$_x$CuO$_4$.
The room temperature resistivity of optimal doped samples of
Pr$_{2-x}$Ce$_x$CuO$_4$ and Nd$_{2-x}$Ce$_x$CuO$_4$ films were
0.26 and 0.23\,m$\Omega$cm, respectively, whereas for undoped
samples the respective values were 5.6 and 8.9\,m$\Omega$cm.

\begin{figure}[t]
\centering{%
\includegraphics[width=0.85\columnwidth,clip=]{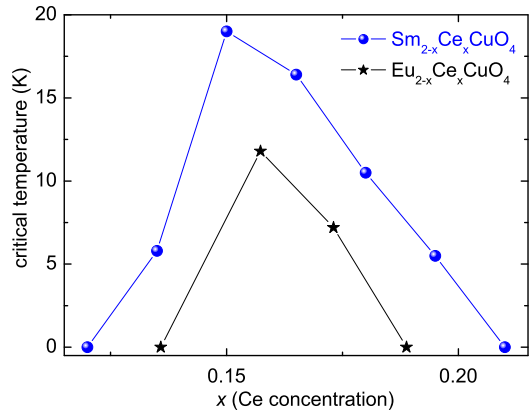}}
 \caption{(Color online) Superconductivity phase diagrams of Sm$_{2-x}$Ce$_x$CuO$_4$ and
 Eu$_{2-x}$Ce$_x$CuO$_4$.}\label{Fig:SCCO}
\end{figure}

{\em Sm$_{2-x}$Ce$_x$CuO$_4$}: The ionic radius of eight-fold
coordinated Sm$^{3+}$ is $1.079$\,{\AA}. From Fig.~\ref{Fig:SCCO}
one sees that the maximal $T_{\text{C}}$ of 19\,K is obtained for
$x=0.150$, in consistency with earlier bulk data \cite{Nagata:02}.
The room-temperature and low-temperature resistivity values show
also minima around $x\approx0.15$. The resistivity range at room
temperature for Sm$_{2-x}$Ce$_x$CuO$_4$ thin films lies between
$0.2$\,m$\Omega$cm and 100\,m$\Omega$cm for optimal doped and
undoped samples, respectively. Superconductivity occurs in the
range of approximately $x\approx0.13$ to $x\approx0.20$. Bulk
superconductivity in Sm$_{2-x}$Ce$_x$CuO$_4$ has been reported by
several authors \cite{Early:93,Prozorov:04}.

{\em Eu$_{2-x}$Ce$_x$CuO$_4$}: The right hand side neighbor of Sm in
the periodic table is Eu. Shortly after electron doped cuprates had
been discovered, Markert et al.~\cite{Markert:89} reported on
superconductivity in Eu$_{2-x}$Ce$_x$CuO$_4$, however, so far no
results on thin films have been reported. The ionic radius of
eight-fold coordinated Eu$^{3+}$ is $1.066$\,{\AA}. The
superconducting region is very limited between $0.14<x<0.19$.
Maximal $T_{\text{C}}$ of 12\,K occurs at $x=0.16$. The phase
diagram is shown in Fig.~\ref{Fig:SCCO}. The resistivity range at
room temperature for Eu$_{2-x}$Ce$_x$CuO$_4$ thin films lies between
$0.6$\,m$\Omega$cm and 140\,m$\Omega$cm for optimal doped and
undoped samples, respectively.

The combined results of our thin-film study of
$R_{2-x}$Ce$_x$CuO$_4$ (with $R=$\,La, Pr, Nd, Sm, Eu, and Gd)
show that the superconductivity phase diagrams depend
systematically on the rare earth ionic radius of the compound.
Superconductivity can only be observed above a threshold ionic
size of about 1.053\,{\AA} of Gd$^{3+}$(or tolerance factor
$>0.83$ \cite{Bringley:90}). The superconducting phase region
expands significantly with $LN^{3+}$ ionic radii, i.e., for
$LN$~=~La the superconducting region spans from $x=0.05$ to
$x=0.22$ whereas $LN$~=~Eu, it is from $x=0.14$ to $x=0.19$.
Correspondingly, it is also observed that the maximal $T_{c}$
increases with ionic radii (for $LN$~=~La -- 31\,K and $LN$~=~Eu
-- 12\,K) and it occurs at considerably lower doping (i.e., for
$LN$~=~La at $x=0.09$ whereas $LN$~=~Eu at $x=0.16$). In other
words, with increasing ionic radius, not only the superconducting
phase region expands, but also the maximal $T_{\text{C}}$
increases, and the doping level where it is observed is shifted
towards lower doping. The end-point is set by
La$_{2-x}$Ce$_x$CuO$_4$ which hits at low doping ($x\lesssim0.05$)
the instability line where the structural transition from $T'-$
into $T$-structure occurs. The breakdown of superconductivity in
La$_{2-x}$Ce$_x$CuO$_4$ around $x\approx0.05$ therefore is clearly
due to this structural phase transition. As underdoped samples
below $x\approx0.05$ are not systematically available, the
experimental evaluation of the phase competition between
superconductivity and antiferromagnetism is elusive. The important
issue is clearly how to compare electron and hole doped compounds
experimentally. Based on our results, it is suggestive that this
comparison should be made in the system La$_{2-x}$$X$$_x$CuO$_4$
with $X=$ Sr and Ce. The caveat here is, that one is still dealing
with a different crystal structure.

\begin{figure}[t]
\centering{%
\includegraphics[width=0.85\columnwidth,clip=]{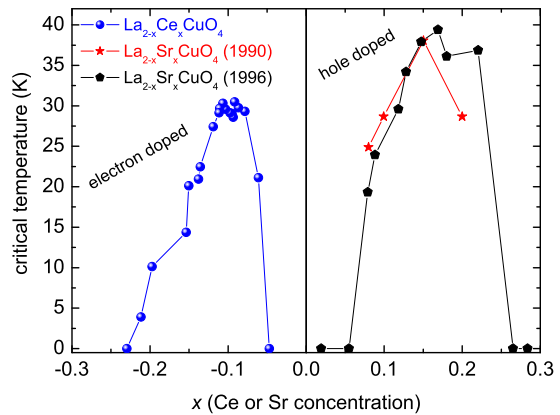}}
 \caption{(Color online) Direct comparison of La-based cuprates with respect to
 electron and hole doping: Superconductivity phase diagrams of
 La$_{2-x}$Ce$_x$CuO$_4$(thin film data)
 and La$_{2-x}$Sr$_x$CuO$_4$ (bulk data \cite{Luke:90,Boebinger:96}).
 }\label{Fig:LXCO}
\end{figure}

The key result of this paper is summarized in Fig.~\ref{Fig:LXCO}.
Here, we compare directly La$_{2-x}$Ce$_x$CuO$_4$ and
La$_{2-x}$Sr$_x$CuO$_4$. The La$_{2-x}$Sr$_x$CuO$_4$ data are
taken from literature and have been obtained from bulk materials
(i.e., no effects of substrate strain are considered that can
increase $T_{\text{C}}$). The comparison of hole and electron
doped cuprates made in Fig.~\ref{Fig:LXCO} is the most direct
possible with available data. The width of the superconducting
phase region is very similar for hole and electron doping. The
electron doped side of the phase diagram even extends to slightly
lower doping. The absolute value of $T_{\text{C}}$ is about 10\,K
(or 33\%) higher in the hole doped case. Surprisingly, the doping
where the maximal $T_{\text{C}}$ occurs at the electron doped side
is shifted to below $x=0.1$ for La$_{2-x}$Ce$_x$CuO$_4$, which is
a reduction by about one third compared to hole doping. There is
no such thing as a distinguished intrinsic doping value of about
$x\approx0.15$ where maximal superconductivity occurs in cuprate
high-temperature superconductors. In most Hubbard model
calculations (see for example recent variational cluster
perturbation theory in the $t-t'-t''-U$ Hubbard model
\cite{Senechal:05} and also recent quantum Monte Carlo simulations
\cite{Aichhorn:06}), the phase competition between $d$-wave
superconductivity and antiferomagnetism comes out like
experimentally observed for the unspecific phase diagram of
Nd$_{2-x}$Ce$_x$CuO$_4$ and Pr$_{2-x}$Ce$_x$CuO$_4$, i.e., for
electron doping the antiferromagnetic phase persists to higher
doping before superconductivity sets in, and the superconducting
region is narrower compared to the hole doping case. With respect
to the superconducting part of the phase diagram, our experimental
results here show that these properties are not intrinsic or
general valid for electron doped cuprates. From our thin film
experiments, we cannot establish the antiferromagnetic region of
the phase diagram. It is well possible that in a certain doping
range, both order parameters coexist \cite{Alff:03,Motoyama:07}.
It remains a theoretical task to understand the presented
superconductivity phase diagrams of electron doped cuprates. Note
that the phase diagram of the electron doped infinite layer
compound Sr$_{1-x}$La$_x$CuO$_2$ where copper is four-fold
coordinated as is the case in the $T'$-structure, shows maximal
superconductivity also around $x\approx0.1$ \cite{Karimoto:01}.

In summary, we have shown by a consistent study of MBE-grown cuprate
thin films, that the superconductivity phase diagrams of the
electron doped cuprates depend strongly on the rare earth ionic
radius. The most suited compound for a comparison of hole and
electron doped cuprates so far is La$_{2-x}$Ce$_x$CuO$_4$. The
extended superconductivity range and the occurance of maximal
$T_{\text{C}}$ at considerably reduced doping indicates  even
slightly higher superconductivity phase stability for the electron
doped side of the phase diagram.

The authors acknowledge the support by the DFG (FOR 538) and
fruitful discussions with Ch. Bernhard, D. Manske, P. Yordanov,
and H. Yamamoto.

\end{document}